# The structure of suspended graphene sheets

Jannik C. Meyer[1], A. K. Geim[2], M. I. Katsnelson[3], K. S. Novoselov[2], T. J. Booth[2], S. Roth[1]

[1]*Max Planck Institute for Solid State Research, Heisenbergstr. 1, 70569 Stuttgart, Germany*

[2]*Manchester Centre for Mesoscience and Nanotechnology, University of Manchester, Oxford Road, Manchester M13 9PL, United Kingdom*

[3]*Institute for Molecules and Materials, Radboud University of Nijmegen, Toernooiveld 1, 6525 ED Nijmegen, The Netherlands*

**The recent discovery of graphene has sparked significant interest, which has so far been focused on the peculiar electronic structure of this material, in which charge carriers mimic massless relativistic particle [1-3]. However, the structure of graphene – a single layer of carbon atoms densely packed in a honeycomb crystal lattice – is also puzzling. On the one hand, graphene appears to be a strictly two-dimensional (2D) material and exhibits such a high crystal quality that electrons can travel submicron distances without scattering. On the other hand, perfect 2D crystals cannot exist in the free state, according to both theory and experiment [4-9]. This is often reconciled by the fact that all graphene structures studied so far were an integral part of larger 3D structures, either supported by a bulk substrate or embedded in a 3D matrix [1-3,9-12]. Here we report individual graphene sheets freely suspended on a microfabricated scaffold in vacuum or air. These membranes are only one atom thick and still display a long-range crystalline order. However, our studies by transmission electron microscopy (TEM) have revealed that suspended graphene sheets are not perfectly flat but exhibit intrinsic microscopic roughening such that the surface normal varies by several degrees and out-of-plane deformations reach 1 nm. The atomically-thin single-crystal membranes offer an ample scope for fundamental research and new technologies whereas the observed corrugations in the third dimension may shed light on subtle reasons behind the stability of 2D crystals [13-15].**

The question whether a strictly 2D crystal can exist was first raised theoretically more than 70 years ago by Peierls [4,5] and Landau [6,7]. They showed that, in the standard harmonic approximation [16], thermal fluctuations should destroy long range order, essentially resulting in melting of a 2D lattice at any finite temperature. Furthermore, Mermin and Wagner proved that a magnetic long-range order could not exist in one and two dimensions [17] and, later, extended the proof to the crystalline order in 2D [8]. Importantly, numerous experiments on thin films have been in accord with the theory, showing that below a certain thickness, typically of many dozens atomic layers, the films become thermodynamically unstable (segregate into islands or decompose) unless they constitute an inherent part of a 3D system (for example, grown on top of a bulk crystal with a matching lattice) [18,19,20]. However, although the theory does not allow perfect crystals in 2D space, it does not forbid nearly perfect 2D crystals in 3D space. Indeed, a detailed analysis of the 2D crystal problem beyond the harmonic approximation has led to a conclusion [13-15] that the interaction between bending and stretching long-wavelength phonons can in principle stabilize atomically thin membranes through their deformation in the third direction [15]. Our experiments described below show that graphene crystals can exist without a substrate, i.e. as freely suspended, and exhibit random elastic deformations involving all three dimensions.

The preparation of graphene membranes used in this study is described in Supplementary Information. Briefly, we used the established procedures [9] of micromechanical cleavage and identification of graphene, followed by electron-beam lithography and a number of etching steps, which allowed us to obtain graphene crystallites attached to a micron-sized metallic scaffold. Figure 1 shows the bright-field TEM image of one of our samples. Central parts of the prepared membranes normally appear on TEM images as homogeneous and featureless regions whereas the membranes' edges tend to scroll (Fig. 1). Also, we often observed folded regions where a graphene sheet became partly detached from the scaffold during microfabrication (right part of Fig. 1). Such folds provide a clear TEM signature for the number of graphene layers. A folded graphene sheet is locally parallel to the electron beam and, for single-layer graphene, a fold exhibits only one dark line (Fig. 2a), similar to TEM images from one half of a single-walled carbon nanotube. For comparison, Fig. 2b shows a folded edge of bilayer graphene, which exhibits two dark lines, as in the case of double-walled nanotubes. One has to be careful, however, because scrolls and multiple folds can give rise to any number of dark lines even for single-layer graphene, as indeed observed experimentally.

In addition, we could directly distinguish between single-layer graphene and thicker samples by analysing nanobeam electron diffraction patterns from their flat areas as a function of incidence angle. This procedure effectively allowed us to probe the whole 3D reciprocal space. Figure 2 shows examples of diffraction patterns at three tilt angles for the graphene membrane of Fig. 1. As expected, there are two dominant reflections corresponding to periodicities of 2.13 Å and 1.23 Å, and weak higher-order peaks. The key for identification of single-layer graphene is that its reciprocal space (Fig. 3) has only the zero order Laue zone and, therefore, no dimming of the diffraction peaks should occur at any angle, in contrast to the behaviour of crystal lattices extended in the third direction. This is exactly the behaviour observed experimentally. Fig. 2f plots the total intensity for diffraction peaks (0-110) and (1-210) as a function of tilt angle for single-layer graphene. One can see that changes in the total intensity are relatively small and, importantly, there are no minima, in agreement with our numerical simulations (see caption to Fig. 2). For comparison, Fig. 2g shows the corresponding behaviour for bilayer graphene, where the total intensities vary so strongly that the same peaks become completely suppressed at some angles and the underlying 6-fold symmetry remains undisturbed only for normal incidence. The diffraction analysis also yields that our bilayer membranes retained the Bernal (AB) stacking of bulk graphite, in contrast to the AAA... stacking reported in "carbon nanofilms" [25]. Note that, independently of stacking order, the weak monotonic variation of diffraction intensities with tilt angle is a unique signature of single-layer graphene and can be employed for its unambiguous identification in TEM.

Notwithstanding the overall agreement, there is one feature in the observed diffraction patterns, which strongly disagrees with our numerical simulations and, more generally, with the standard diffraction behaviour in 3D crystals [21,22]. Indeed, one can readily see that the diffraction peaks in Fig. 2 become broader with increasing tilt angle and the blurring is much stronger for those peaks that are further away from the tilt axis. This broadening is a very distinctive feature of single-layer graphene. It becomes notably weaker in bilayer samples and completely disappears for multilayer graphene. From a theory point of view, the broadening is completely unexpected. To emphasize this, we note that, for example, thermal vibrations can only reduce the intensity of diffraction peaks (Debye-Waller factor) but do not lead to their broadening [21,23].

Figure 3 explains how the observed broadening explicitly yields that graphene sheets are not flat within the submicron area of the electron beam. The full 3D Fourier transform of a flat graphene crystal (Fig. 3a) consists of a set of rods perpendicular to the plane of the reciprocal hexagonal lattice (Fig. 3c). Each diffraction pattern is then a two-dimensional slice (given by a section of the Ewald sphere) through this 3D space. In particular, this picture suggests that the intensity of diffraction peaks should vary without any singularities (monotonically) with changing tilt angle and the hexagonal symmetry is preserved for any tilt, as already discussed above. The increasing broadening of diffraction peaks without changes in their total intensity implies that the rods wander around their average direction (see Fig. 3d). This corresponds to a slightly uneven sheet (Fig. 3b) so that the diffraction pattern effectively comes from an ensemble of small flat 2D crystallites with different orientations with respect to the average plane. Fig. 3e illustrates that such roughness results in sharp diffraction peaks for normal incidence but the peaks rapidly become wider with increasing tilt angle. This model also yields that their total intensity should be practically independent of membrane's roughness and can be described by the angle dependence for a flat sheet, which is consistent with our simulations in Fig. 2f.

For quantitative analysis, Figures 3f,g show the detailed evolution of broadening of the diffraction peaks with changing incidence angle. One can see that the peaks' widths increase linearly with tilt and, also, proportionally to the peaks' position in reciprocal space, in quantitative agreement with our simulations for corrugated graphene. The width of the cones or the linear slopes in Figs. 3f,g provide a direct measure of the membrane's roughness. For different single-layer membranes, we found cone angles between 8° and 11°, that is, the surface normal deviated from its mean direction on average by ±5°. For two-layer membranes, this value was found to be twice smaller, i.e. ≈2° (Fig. 3g). It is important to note that the diffraction peaks broaden isotropically (see Figs. 2 c-e). This means that the surface normal in real space wanders in all directions, and the observed waviness is *omni-directional*. Otherwise, if a graphene membrane were curved only in one direction, diffraction peaks would spread into a line indicating the direction of curling. An absolutely incompressible sheet can only be curved in one but not two directions, and the isotropic waviness unambiguously implies local deformations of graphene. The curvature of 5° yields a local strain of up to 1%, which is large but sustainable without plastic deformation and generation of defects [26-28].

To estimate the spatial extent $L$ of the found corrugations, let us start with two observations. First, $L$ cannot be drastically (more than a few times) smaller than the coherence length of the diffracted electrons. Otherwise, we would expect sharp peaks and also much stronger deviations between the experimental and calculated intensities in Fig. 2f. The coherence length is estimated to be ≈10nm, so that the corrugations must have a mesoscopic (several nm) rather than atomic scale. Second, the smooth Gaussian shape of the diffraction peaks requires a large number $N$ of different orientations within the submicron illuminated area, which provides us with the upper limit for $L$. A minimalist assumption of $N$=100 necessitates $L$≤25 nm. These qualitative considerations are in agreement with our simulated diffraction patterns for corrugated graphene sheets (see Supplementary Information). From the known curvature and size $L$ of the corrugations, we estimate their height as ≈1 nm.

The above order-of-magnitude estimates are also strongly supported by atomic resolution TEM imaging of our membranes. Unfortunately, for single-layer graphene, such smooth waviness could not be visualized because diffraction intensities vary little with tilt angle, as discussed earlier (Fig. 2f), and no additional contrast due to corrugations could be expected or, in fact, observed. On the other hand, the visibility of the hexagonal lattice for two and more layers strongly depends on their tilt angle (Fig. 2g) and, accordingly, surface undulations of few-layer graphene can be expected to result in areas of different brightness. Such areas are clearly seen in Fig. 4 and have a characteristic size of a few nm, which is somewhat smaller than the above estimate for $L$ in single-layer graphene where the ripples could also be larger laterally. Importantly, atomic-resolution images show that the corrugations are static, as otherwise, changes during the exposure would lead to blurring and disappearance of the additional contrast.

As mentioned in the introduction, perfect 2D atomic crystals cannot exist, unless they are of a limited size or contain many crystal defects [7,8]. The observed microscopic corrugations of 2D graphene in the third dimension provide an alternative and rather unexpected route to reconcile the high quality of graphene with its thermodynamic stability. On the one hand, the fact that the microscopic roughness is reproducible for different positions on membranes and for different samples, becomes notably smaller for bilayer graphene and disappears for thicker membranes proves that the corrugations are intrinsic to graphene membranes. On the other hand, theoretical investigations of 2D membranes have predicted their thermodynamic stability through static microscopic crumpling involving either bending or buckling [13-15]. The buckling mechanism requires the generation of dislocations [15] that are neither observed nor expected in our case of relatively small (micron-sized) membranes and strong inter-atomic bonds. On the other hand, the bending scenario assumes no defects and only requires out-of-plane deformations involving a significant elastic strain. The latter is in qualitative agreement with our observations but further experimental and theoretical studies are required to clarify the detailed mechanism of the corrugations in graphene. It is also important to note that the presence of elastic corrugations is consistent with high mobility of charge carriers in graphene [1-3] and can even explain some of its unusual transport characteristics such as the suppression of weak localization [29].

In conclusion, free-hanging graphene is the thinnest conceivable object and offers many exciting directions for future research. The observed microscopic roughening seems to be essential for the structural stability of 2D membranes and indicates that their mechanical, electronic, optical and other properties can be equally extraordinary. 2D crystal membranes also promise such tantalizing applications as virtually transparent substrates for high-resolution electron microscopy and sieving of atoms and small molecules through the atomic-size benzene rings and can be considered for any other technology in which ultra-thin, transparent and robust substrates offer an advantage (for example, nanomechanical devices).

**Acknowledgements**. We thank B. Freitag and D. Beamer of FEI for providing access to their in-house TEM Titan. This work was supported by the EU project CANAPE, the EPSRC (UK) and the Royal Society. M.I.K. acknowledges financial support from FOM (Netherlands).

Correspondence and requests for material should be addressed to J.C.M. (email @ jannikmeyer.de) and A.K.G. (geim @ man.ac.uk)

**Figures**

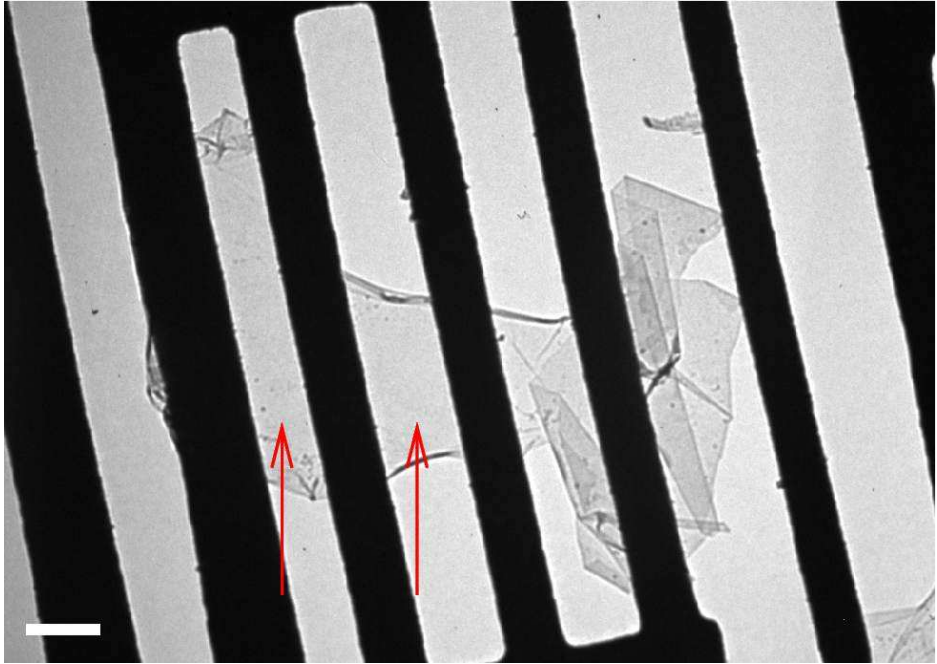

Figure 1: **Suspended graphene membrane.** Bright-field TEM image of a suspended graphene membrane. Its central part (homogeneous and featureless region indicated by arrows) is single-layer graphene. Electron diffraction images from different areas of the flake show that it is a single crystal without domains. Note scrolled top and bottom edges and a strongly folded region on the right. Scale bar 500 nm.

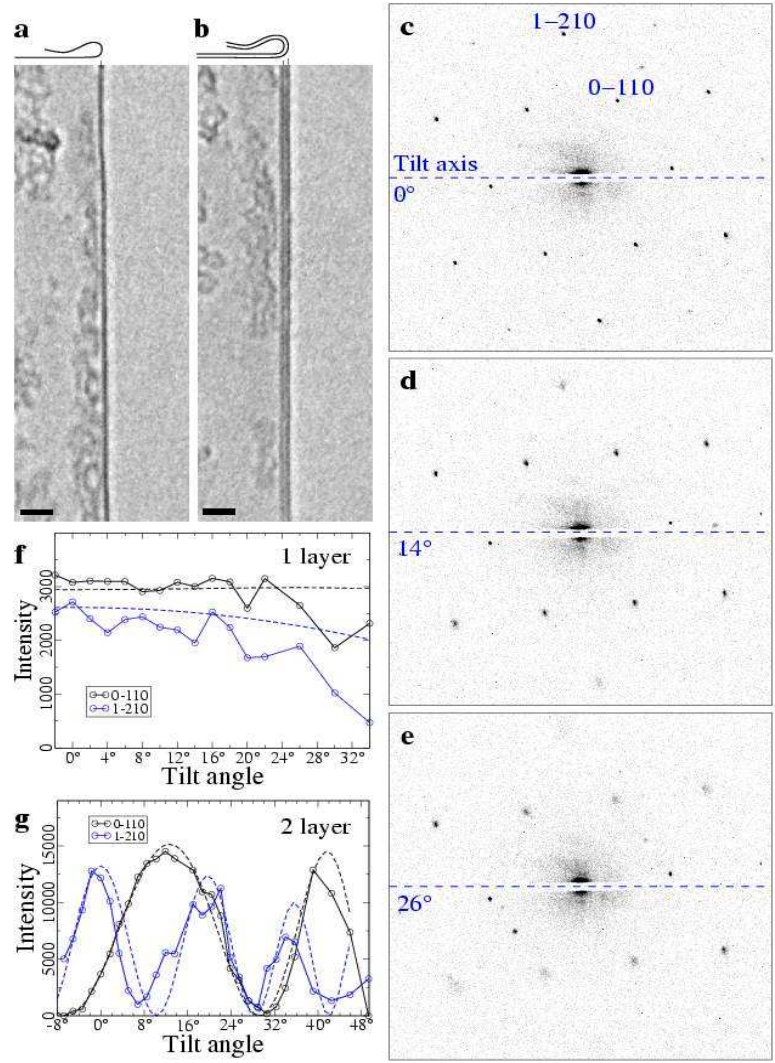

Figure 2: **Transmission electron microscopy of graphene. a** and **b**, TEM images of folded edges for single- and bi- layer graphene, respectively, using a Philips CM200 TEM. Scale bars 2 nm. **c** to **e**, Electron diffraction patterns from a graphene monolayer under incidence angles of 0°, 14° and 26°, respectively. The tilt axis is horizontal. Here we used a Zeiss 912 TEM operated at 60 kV in the Köhler condition with the smallest (5 μm) condenser aperture. This allowed us to obtain a small, practically parallel beam with an illumination angle of 0.16 mrad and an illumination area of 250 nm in diameter only. The diffraction patterns were recorded on CCD for further quantitative analysis. Note that the peaks become broader with increasing tilt, and this effect is strongest for peaks further away from the tilt axis. To label equivalent Bragg reflections, we use the Miller-Bravais indices (*hkil*) for graphite so that the innermost hexagon and the next one correspond to indices (0-110) (2.13 Å spacing) and (1-210) (1.23 Å spacing), respectively. **f**, Total intensity as a function of tilt angle for the peaks marked in panel **c**. To find the intensity values, each of the above Bragg reflections was fitted by a Gaussian distribution for every angle, which yielded the peaks' intensities, positions, heights and widths. The dashed lines are numerical simulations, in which we used a Fourier transform of the projected atomic potentials [21,22,23] and the atomic form factors reported in [24]. **g**, The same analysis and simulations for a two-layer graphene membrane.

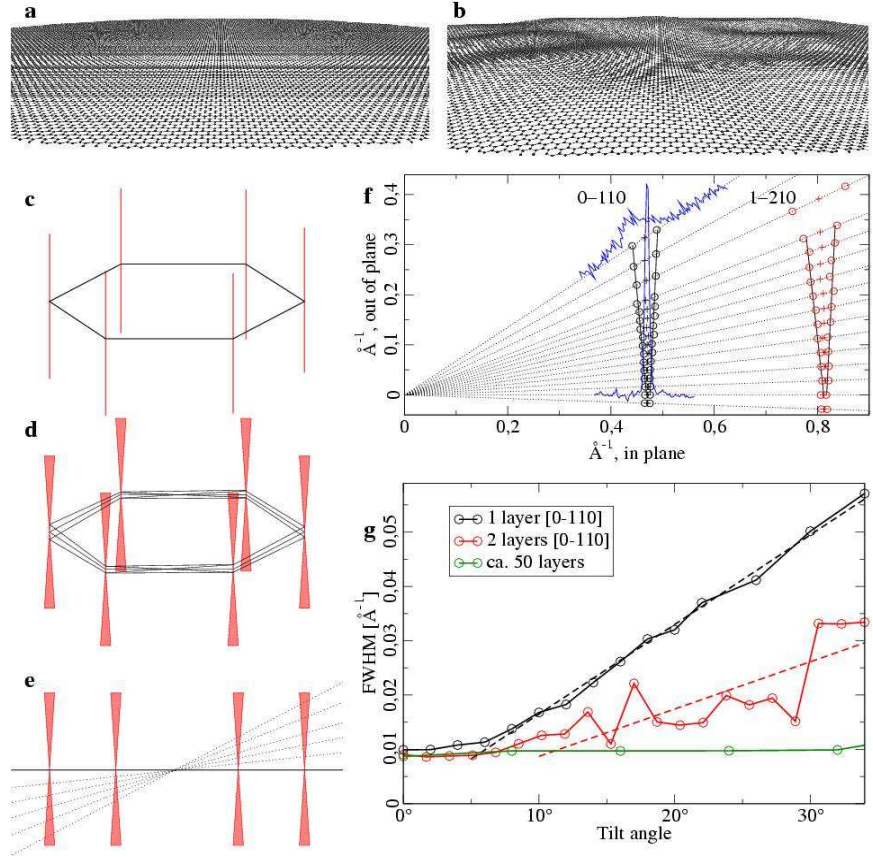

Figure 3: **Microscopically corrugated graphene. a**, Flat graphene crystal in real space (perspective view). **b**, The same for corrugated graphene. The shown roughness imitates quantitatively the one found experimentally. **c**, The reciprocal space for a flat sheet is a set of rods (red) directed perpendicular to the reciprocal lattice of graphene (black hexagon). **d** and **e**, For the corrugated sheet, a superposition of the diffracting beams from microscopic flat areas effectively turns the rods into cone-shaped volumes so that diffraction spots become blurred at large angles (indicated by the dotted lines in **e**) and the effect is more pronounced further away from the tilt axis (compare with Fig. 2). Diffraction patterns obtained at different tilt angles allow one to measure graphene roughness. **f**, Evolution of diffraction peaks with tilt angle in single-layer graphene. The experimental data are presented in such a way that they closely resemble the schematic view in panel **e**. For each tilt angle, the black dotted line represents a cross section for diffraction peaks (0-110) and (1-210). The peaks' centres and full widths at half maxima (FWHM) in reciprocal space are marked by crosses and open circles, respectively. In two cases (0° and 34°), the recorded intensities are shown in full by blue curves. All the intensity curves could be well fitted by the Gaussian shape. The solid black lines show that the width of the diffraction spots reproduces the conical broadening suggested by our model (panels **d** and **e**). **g**, FWHM for the (0-110) diffraction peak in single- and bi-layer membranes and thin graphite, as a reference, as a function of tilt angle. The dashed lines are the linear fits yielding the average roughness. The flat region between 0° to 5°, and also for the reference sample, is due to the intrinsic peak width for the microscope at our settings.

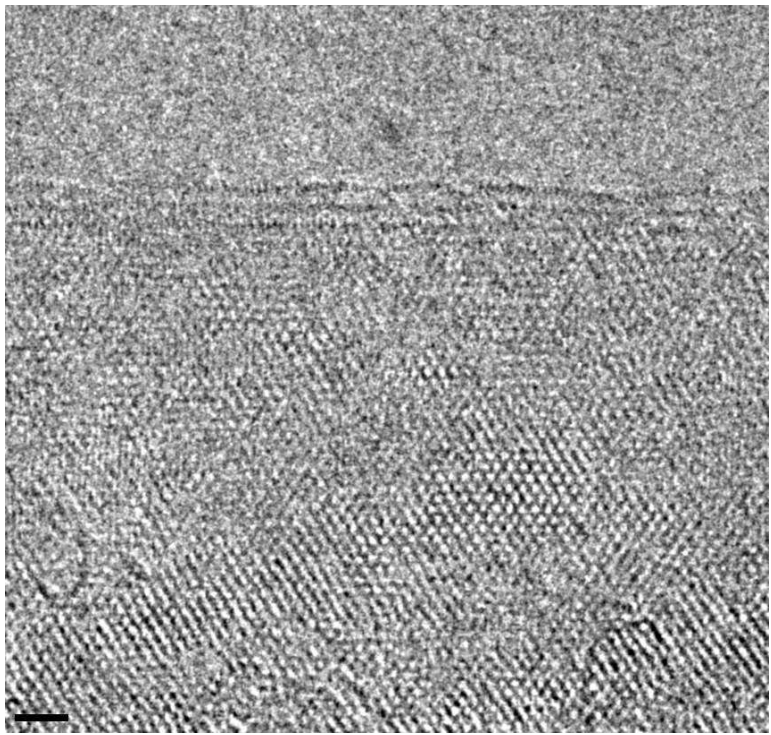

Figure 4: **Atomic resolution imaging of graphene membranes.** TEM image of a few-layer graphene membrane near its edge, where the number of dark lines indicates the thickness of 2 to 4 layers. Because for few-layer graphene the electron contrast depends strongly on incidence angle, relatively small (few degrees) variations in the surface normal become visible. The atomic-resolution imaging was achieved by using FEI Titan at an acceleration voltage of 300 kV. Scale bar 1 nm.

# Supplementary information

**Sample preparation**

To make graphene membranes, we started with graphene flakes prepared on top of an oxidized silicon wafer (300 nm of $SiO_2$) by micromechanical cleavage, as described previously [1,9]. Monolayer flakes were identified by optical microscopy from a subtle shift in colour [1,9] as compared to the empty surface (Fig. S1A) and, if in doubt, this was double-checked by atomic force microscopy [1,9]. A metal grid (3 nm Cr and 100 nm Au) was then deposited on top of a chosen flake by using electron-beam lithography. After that, the substrate was cleaved so that its edge was within 50μm from the chosen flake (Fig. S1B). These samples were put in 15% tetramethylammonium hydroxide at 60°C for several hours, which etched away the bulk Si, undercutting the grid. The etching was monitored through an optical microscope and stopped after a sufficient part of the grid became overhanging (Fig. S1C). The remaining $SiO_2$ layer was removed during 5 minutes in 6% buffered hydrofluoric acid. The samples were then transferred into water, isopropanol, acetone and, finally, liquid carbon dioxide for critical point drying. Fig. S1C,D show the resulting scaffold with the flakes attached underneath it.

**Numerical simulations**

To estimate the spatial size of the microscopic crumpling, we have performed simulations of the electron diffraction patterns that are expected for non-flat graphene sheets in the partially coherent illumination of our TEM. To this end, we took a flat graphene sheet (Fig. S2A) and added to this out-of-plane ($z$) displacements with single-frequency components given by $z(x,y)=A\sin(k_x x+k_y y+B)$ where $A$, $B$, $k_x$, $k_y$ were random parameters. A single-frequency component is shown in Fig. S2B. A large number of such sinusoidal waves were superimposed to obtain a randomly curved sheet, as shown in Fig. S2C. The random parameters were distributed so that desired average for out-of-plane deformation $h$ and for a lateral ripple size $L$ were obtained. For each tilt angle, the projected atomic potentials were then calculated by using a 10 nm area of the sheet (the latter size corresponds to the coherence length of electrons in our experiments). After this, we calculated the Fourier transform of the projected potentials, which yields diffraction patterns from each coherently illuminated area of 10 nm in size. Because experimentally we worked with the smallest possible but still relatively large beam (250 nm in diameter), we calculated diffraction patterns from many coherently illuminated units and added up their intensities.

The described simulation at different tilt angles were carried out for various sizes and heights of the ripples. In all the cases, the average waviness was fixed at 5° (as measured experimentally), which corresponds to ratio $L/h \approx 10$ between ripples' lateral size $L$ and their height $h$. Figures S3 to S5 show examples of the resulting simulations of diffraction patterns for a tilt angle of 26°, which mimics the experimental situation in Fig. 2E of the main paper. If ripples' size is notably smaller than the electron coherence length, the simulated diffraction pattern displays sharp peaks (Fig. S3), in clear disagreement with our experimental observations. In the opposite limit of ripples larger than the coherence length (Fig. S5), diffraction spots become representative of the actual local bending of a graphene sheet. This contradicts to our experiment, where we observe diffraction peaks with a smooth Gaussian shape, independently of the spot position and using different samples. This indicates that in our case the average ripple size should be of the order of the electron coherence length. In this case, the simulated peak broadening (Fig. S4) closely resembles our experimental data in Fig. 2 and 3. Further comparison between the simulations and experiment infers ripples' lateral size of 5 to 10 nm. However, because of experimental uncertainty in the actual coherence length, which may be a factor of 2 larger or smaller than the value of 10 nm used in our simulations, we make a prudent estimate for the typical lateral size of crumpling as between 2 and 20 nm.

Figures

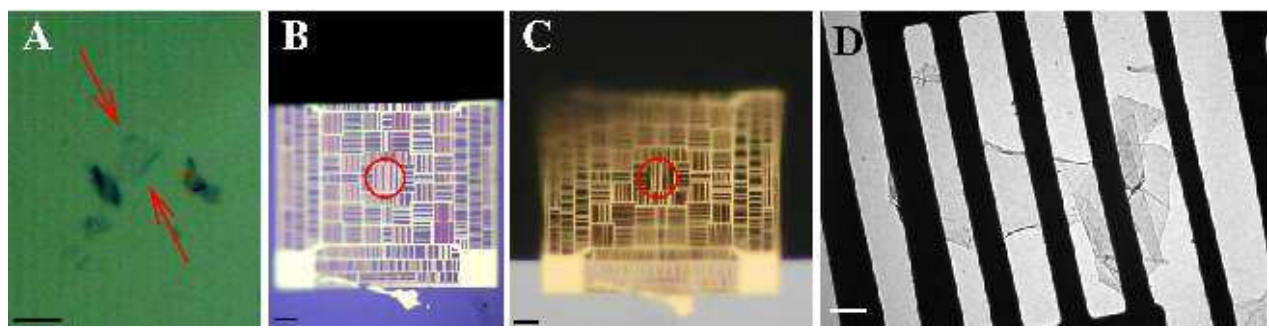

Figure S1: **Preparation of suspended graphene membranes. A,** Optical micrograph of a single-layer graphene sheet (arrows) on an oxidized Si substrate, which is surrounded by a few thicker flakes. **B**, Metal grid deposited on top of the graphene sheet. **C**, Part of the substrate is removed by chemical etching so that the metal grid reaches over the substrate edge. The graphene sheet (marked by red circles in B and C) remains attached to the metal scaffold. **D**, Bright-field TEM image of a suspended graphene membrane. Scale bars: 5 µm (**A**), 10 µm (**B,C**) and 500 nm (**D**).

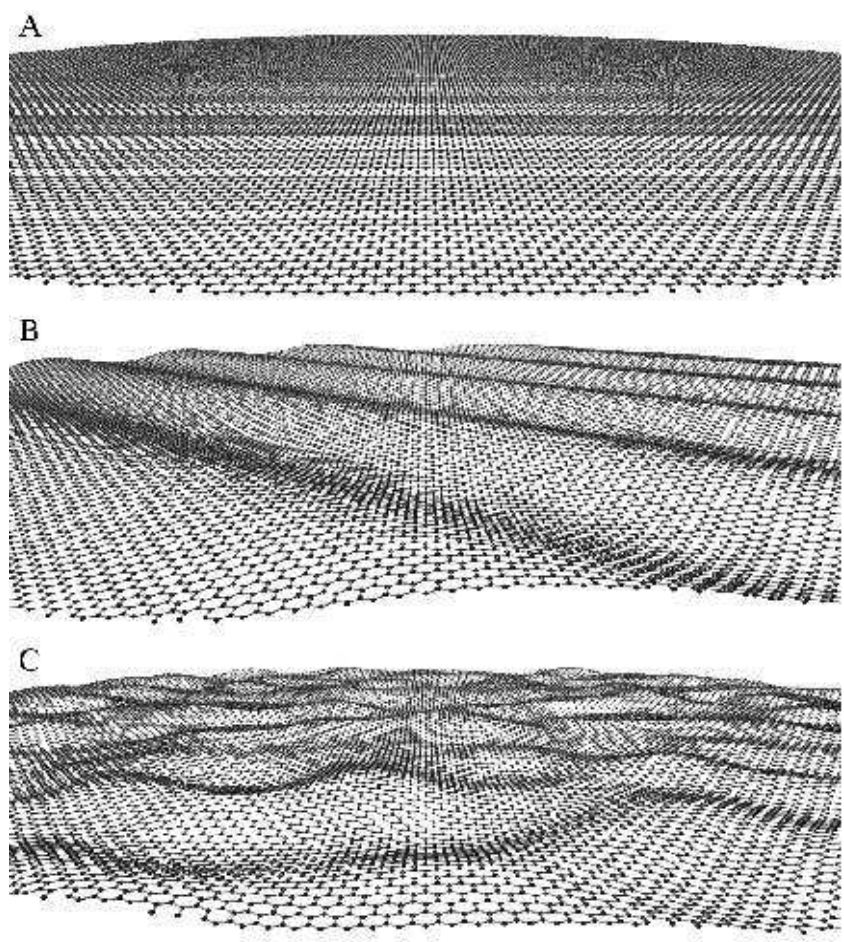

Figure S2: Modelling of crumpled graphene. Starting from a flat sheet (**A**), a number of random waves such as the one shown in **B** are introduced to form the randomly curved membrane (**C**).

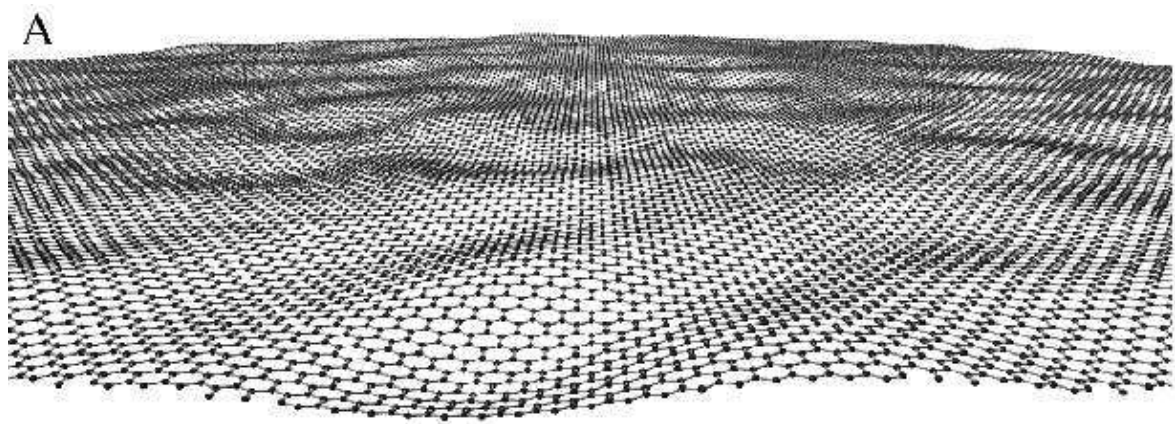

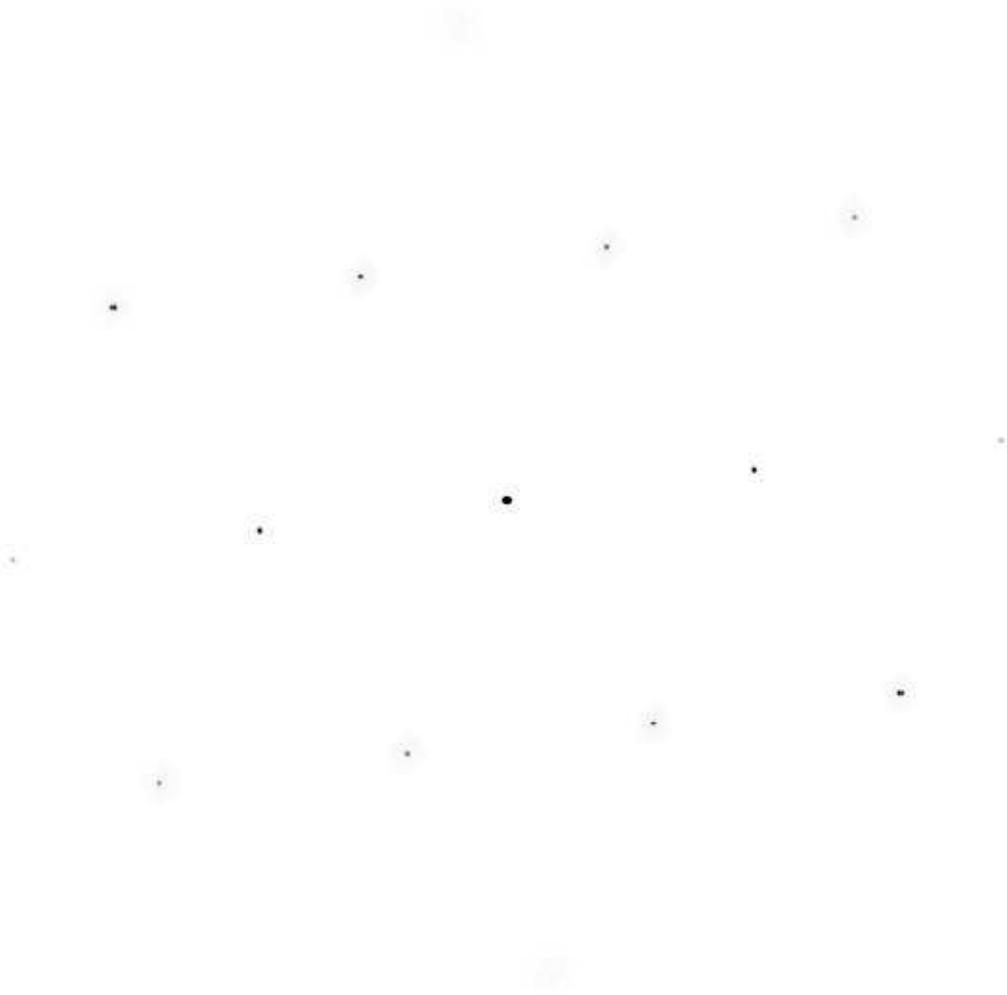

Figure S3: **A**, Graphene sheet with ripples of, typically, 0.2 nm in height $h$ and 2 nm in lateral size $L$. **B**, Simulated diffraction pattern. Such ripples are too small to explain our observations.

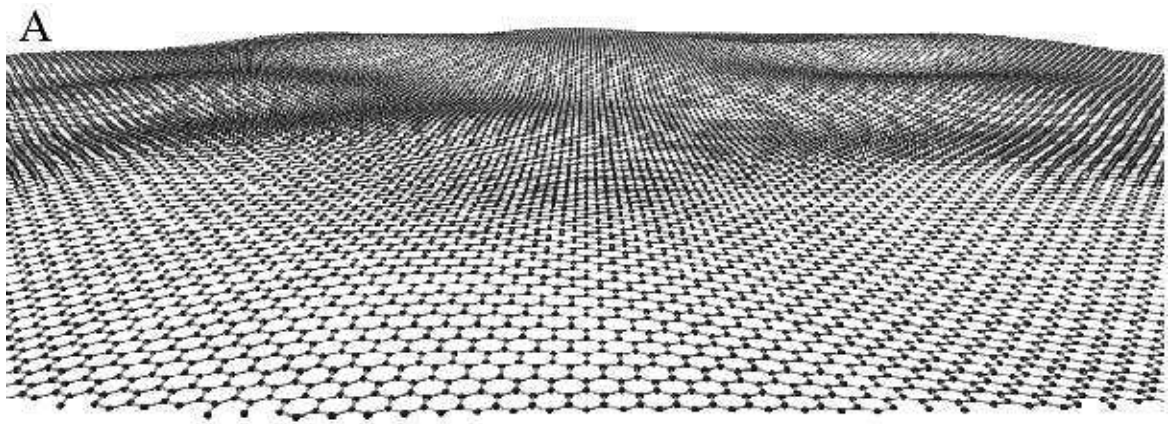

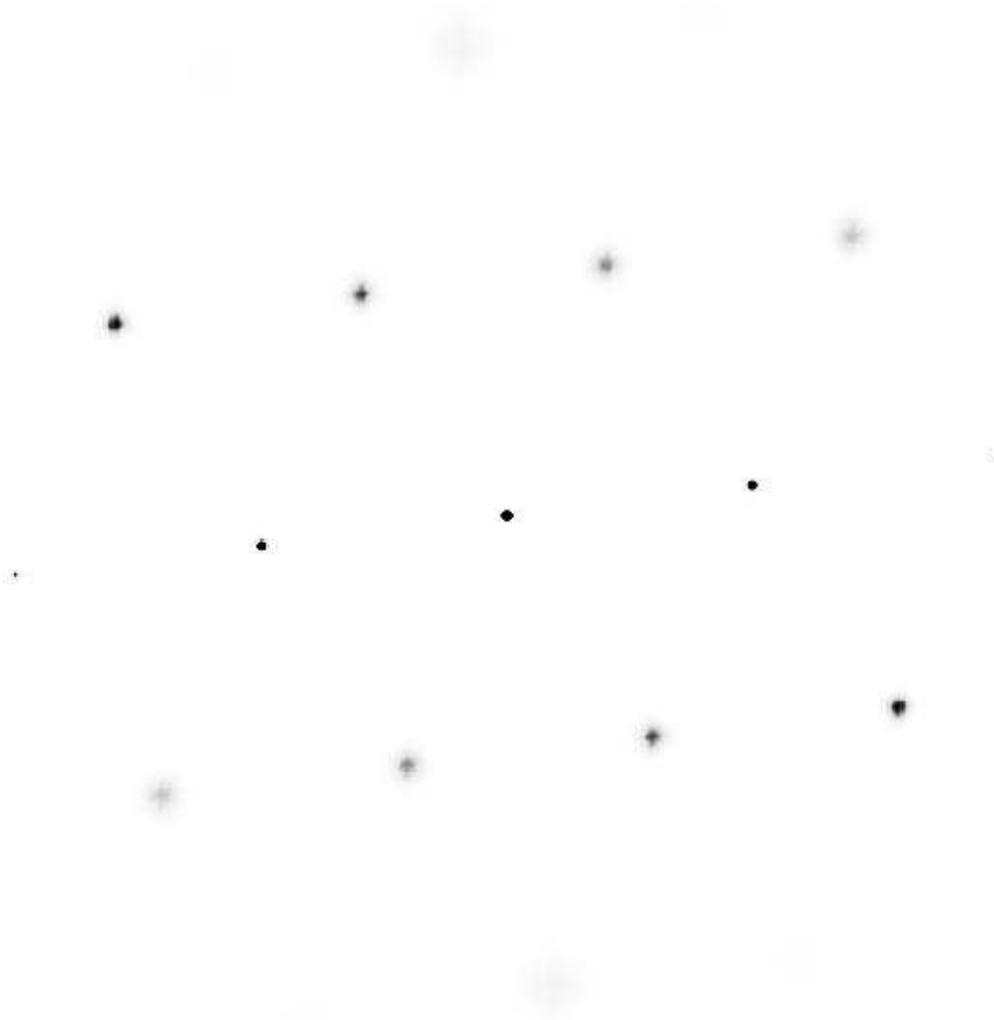

Figure S4: Stronger crumpling. **A**, Ripples are 0.5 nm in height and their typical size is 5 nm laterally. Ratio L/h is the same as in Fig. S3. **B**, Calculated diffraction patterns agree well with our experimental data.

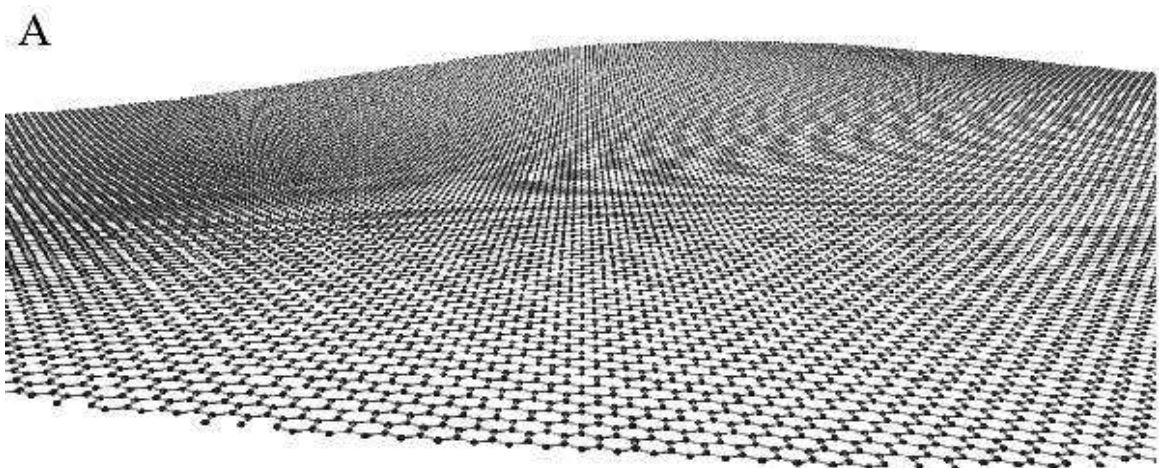

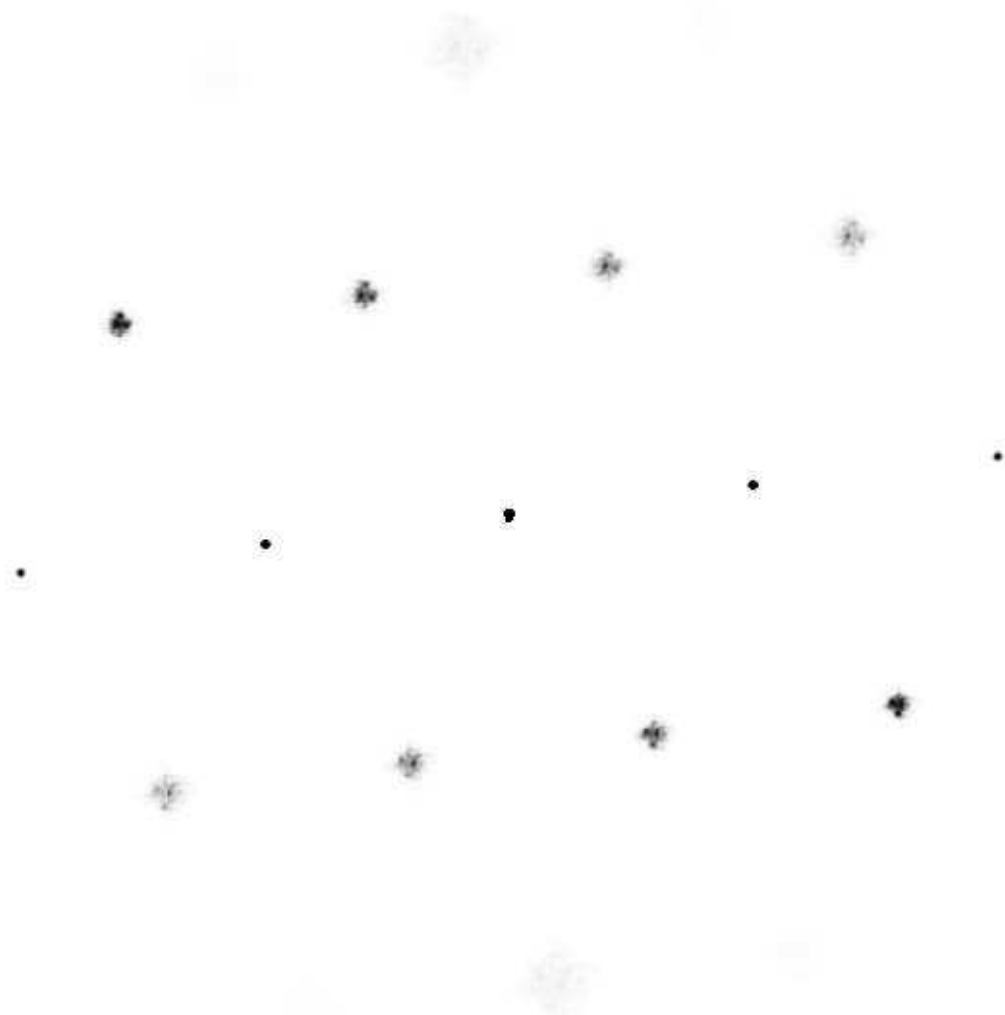

Figure S5: Ripples 2 nm high and 20 nm wide (**A**). Such large-scale crumpling leads to a non-Gaussian intensity distribution in the broadened diffraction peaks (**B**) which start reflecting specific distortions of a graphene sheet within the 250 nm diameter diffracting beam. Ripples of this size cannot be dominant in our membranes.

14